\documentstyle[12pt,thmsa,sw20lart]{article}

\input tcilatex
\QQQ{Language}{
British English
}

\begin{document}

\section{Introduction}

The conceptual problems of quantum mechanics still attract much attention.
Among the most famous and illustrative of these problems are the
Schr\"odinger cat paradox[1] and the problem of wave function reduction[2].
The ruling Copenhagen interpretation of quantum mechanics does not address
them to the satisfaction of all. This has led to the development of variant
interpretations of quantum mechanics. In this paper, we explain how the
interpretation of quantum mechanics due to Land\'e deals with these problems
in a natural and satisfying way.

We briefly outline the cat paradox and the collapse of the wave function.
Both of them arise from the superposability of quantum states. Let the
eigenfunctions $\psi _1(x)$ and $\psi _2(x)$ be solutions of the
Schr\"odinger equation corresponding to the energy eigenvalues $E_1$ and $%
E_2 $. The functions $\psi _1(x)$ and $\psi _2(x)$ are states of the system;
so also is their superposition 
\begin{equation}
\Psi (x)=a_1\psi _1(x)+a_2\psi _2(x)  \label{on1}
\end{equation}
where $a_1$ and $a_2$ are complex constants. If the energy is measured while
the system is in the state $\Psi (x)$, either $E_1$ or $E_2$ or is obtained;
the respective probabilities of obtaining these eigenvalues are $\left|
a_1\right| ^2$ and $\left| a_2\right| ^2.$ If in such a measurement the
value $E_1$ is obtained, then another measurement of the energy will yield $%
E_1$ with certainty, because the first measurement of the energy threw the
system into the state $\psi _1(x).$ The fact that a measurement of the
energy when the system is in the state $\Psi (x)$ can give either $E_1$ or $%
E_2$or shows that in that state, the system is to be considered as being in
a superposition of the states $\psi _1(x)$ and $\psi _2(x)$. The act of
measuring the energy actualizes one of the eigenvalues of the energy. In
this example, the process whereby the system state changes from the
superposition $\Psi (x)$ to $\psi _1(x)$ or $\psi _2(x)$ constitutes
collapse or reduction of the wave function.

Superposition states are a general feature of quantum theory and can be
formed with the eigenfunctions of any observable, not just the energy. A
hypothetical observable in which the eigenvalues are the states ''dead'' and
''alive'' of a cat illustrates very strikingly the implications of
superposition states. With this observable, the state

\begin{equation}
\Psi =a_{\text{dead}}\psi _{\text{dead}}+a_{\text{alive}}\psi _{\text{alive}}
\label{tw2}
\end{equation}
is possible. Since this is a superposition state, the cat is neither dead
nor alive until an experiment is performed that reduces the system wave
function to that of the state ''dead'' or ''alive''. The conclusion that a
macroscopic entity like a cat can be in a superposition of two states is the
essence of the cat paradox.

The conceptual problems arising from the collapse of the wave function and
from the cat paradox are well known. Next we show how the Land\'e
interpretation of quantum mechanics disposes of them.

\section{Background Theory}

The Land\'e interpretation of quantum mechanics was outlined in a series of
books and papers[3-6]. Here we highlight those aspects of it needed to
discuss the cat paradox and wave function collapse.

According to the Land\'e approach, nature is inherently indeterministic and
this fact must be ready-built into quantum theory[4]. In treating
measurements, we must therefore be reconciled to the fact that all we can
hope to do with theory is to predict the probabilities of various possible
outcomes of an experiment. If a quantum system contains several observables,
then we can hope to use theory to determine the probabilities of obtaining
given values of a particular observable after having measured and obtained
specific values of another.

Let a quantum system possess the three observables $A$, $B$ and $C$ with
respective eigenvalue spectra $A_1$, $A_2$,...$A_N$, $B_1$, $B_2$,...$B_N$
and $C_1$, $C_2$, ...$C_N$. A measurement of $B$ from a state defined by the
eigenvalue $A_i$ of the observable $A$ can yield any of the eigenvalues $B_j$
of $B$ with probabilities determined by the probability amplitudes $\chi
(A_i,B_j).$ Similarly, a measurement of $C$ is described by the probability
amplitudes $\psi (A_i,C_j).$ On the other hand, a measurement of $C$ from
the state of $B$ defined by the eigenvalue $B_i$ yields any of the values $%
C_j$ with probabilities given by the probability amplitudes $\phi (B_i,C_j).$
The probabilities for measurements in the opposite direction are given by
the same probability amplitudes. Thus, for example, 
\begin{equation}
\psi (A_i,C_j)=\psi ^{*}(C_j,A_i)  \label{th3}
\end{equation}
This means that the matrix formed from the probability amplitudes is
Hermitian.

As would be expected of quantities pertaining to one system, the three sets
of probability amplitudes are not independent of each other. The relation
that connects them is[4] 
\begin{equation}
\psi (A_i,C_n)=\dsum\limits_{j=1}^N\chi (A_i,B_j)\phi (B_j,C_n)  \label{fo4}
\end{equation}
with identical relations for the other probability amplitude. This is
evidently the law of probability addition, a fundamental axiom of quantum
mechanics. A careful interpretation of this relation is all we need to clear
up the cat paradox and the problem of wave function collapse.

The probability addition law Eq. (\ref{fo4}) is evidently the ultimate
origin of the superposition state. In order to obtain the standard
superposition state from this expression, we need only to label the
quantities incompletely. Firstly, since in the treatment of solutions of the
Schr\"odinger equation, $C$ is the position and is continuous, we make the
assignment $C_n=x$. We omit the labelling of the initial state, so that the
left-hand side becomes $\Psi (x).$ Finally, because the initial state is not
properly identified, the structure of the probability amplitudes $\chi
(A_i,B_j)$ is obscured and we can treat them as mere constants $b_j.$ Hence,
the expansion becomes

\begin{equation}
\Psi (x)=\dsum_{j=1}^Nb_j\phi _j(x)  \label{fi5}
\end{equation}

A second crucial feature of the Land\'e approach is the idea that every
probability amplitude connects a well-defined initial state and a
well-defined final state[4]. Wave functions are just probability amplitudes
and this applies to them as well. From this, we deduce that solutions of
eigenvalue equations connect initial states defined by the eigenvalue and
final states defined by the variable in terms of which the differential
operator is defined[7,8]. To use the Schr\"odinger equation as an example,
its eigenfunctions are probability amplitudes corresponding to the energy
eigenvalues as defining the initial states and the position as
characterising the final state.

\section{Analysis of the Paradoxes}

Having understood that a superposition state is just an under-labelled form
of the probability-addition law Eq. (\ref{fo4}), we can proceed to analyse
the cat paradox and the process of wave function collapse.

As explained above, the origin of the cat paradox is the superposability of
quantum states. Because of the suppression of the initial state index $A_i$,
the role of defining the state passes to the index $j$. But since many
values of $j$ appear in the expansion, we are forced to think of the system
as being in a superposition of states of the observable $B$. We are then
left with no choice but to derive the cat paradox and wave function
reduction from the expansion and to grapple with all the conceptual problems
these conclusions engender.

In truth however, the state represented by an expansion is not described by
one of the values of the expansion observable. The state is actually given
by the label $A_i$. Therefore, a ''superposition state'' is not one in which
the system is in some kind of limbo. It is one in which the system is in an
eigenstate of the observable $A$. Our conceptual problems arise mainly from
our failure to label the probability amplitude $\Psi $ with the
initial-state parameter. In reality, the state $\Psi (x)$ is the probability
amplitude $\Psi (A_i,x).$ In this quantity, the main parameters of interest
are the initial-state eigenvalue $A_i$ and the final-state eigenvalue $x$.
If we desire to express this probability amplitude as an expansion in terms
of the probability amplitudes $\chi (A_i,B_j)$ and $\phi (B_j,x)$, this is
for convenience only and does not mean that the system is in a superposition
of the states of the observable $B$.

It is true that the ability to be able to express a probability amplitude in
terms of two other sets of probability amplitudes is a very powerful tool
indeed. We may sometimes not have an obvious means of finding the
probability amplitude $\Psi (A_i,x)$, while we do have the sets of
probability amplitudes $\chi (A_i,B_j)$ and $\phi (B_j,x).$ In that case the
expansion enables us to find $\Psi (A_i,x)$ in terms of the other sets of
probability amplitudes. However, any other two sets of probability
amplitudes could be used for this purpose, provided that one set corresponds
to the initial-state eigenvalue being $A_i$ and the other set corresponds to
the final-state eigenvalue being $x$. Thus, we could equally use the two
sets $\chi ^{\prime }(A_i,D_j)$ and $\phi ^{\prime }(D_j,x),$ where $D$ is
another observable of the system. But whether the third observable is $B$ or 
$D$, the state of the system is determined only by the eigenvalue $A_i$.

Once we have shifted the definition of the state to the parameter $A_i,$ we
are able to assign to the index $j$ its true role. It is the state label for
the probability amplitudes $\phi _j(x)=\phi (B_j,x)$ which describe
measurements of the observable $C$ from states of the observable $B$. But it
is not the state label for the states described by the probability amplitude 
$\Psi (x)=\Psi (A_i,x).$ The appearance of the label $j$ is entirely
artificial since it arises only from our decision to expand the probability
amplitude $\Psi (A_i,x)$ in terms of the states of the observable $B$. We
could have chosen another observable, say $D$, of the system for this
purpose. Also, this expansion is not essential, and is justified only when
used as a tool for determining the expression for $\Psi (A_i,x)$ in the
event that there is no direct means of obtaining it.

The question arises as to the identity of the observable $A,$ and indeed as
to whether a quantum system is always in a well-defined state. The answer to
this is that before any kind of measurement can be performed on a system,
the system has to exist. This means that it contains some property which can
be measured by which this existence is ascertained. The eigenvalues of this
quantity in principle furnish the values $A_i$. Thus indeed a quantum system
can always be thought to be in a well-defined state when we first encounter
it. This state could be thought of as having being brought about by a
measurement. Since the parameter $A$ always exists, it should never be
omitted from the labelling of the probability amplitudes $\Psi (x),$ which
should in reality always be written as $\Psi (A_i,x).$

At this point, it is necessary to recapitulate the interpretation of
eigenfunctions. In the context of quantum theory, an eigenfunction is a
probability amplitude that connects an initial state corresponding to the
eigenvalue and a final state corresponding to the variable represented by
the variable in terms of which the differential operator is cast. Thus, the
solutions of the time-independent Schr\"odinger equation in coordinate space
connect initial states corresponding to the energy eigenvalues and final
states belonging to position eigenvalues. The solutions of the momentum
eigenvalue equation connect initial states belonging to the linear momentum
and final states corresponding to the position.

\section{Resolution of the Paradoxes}

We are now in a position to dispose of the problem of the collapse of the
wave function of the cat paradox. These are seen to result merely from
assigning a false significance to the index $j$ that appears in the
expansion of the primary probability amplitude $\Psi (A_i,x).$

The problem of wave function collapse vanishes as soon as we realise that a
system in the state $\Psi (A_i,x)$ is not in a superposition of states of
the observable $B.$ The system is in the state corresponding to the
eigenvalue $A_i$. The observable $B$ is not of necessary or fundamental
importance to the description of the probability amplitude $\Psi (A_i,x)$.
In cases where the probability amplitude $\Psi (A_i,x)$ is ready known, it
is not necessary to associate it with the observable $B$.

If on the other hand we ask what the probabilities are of obtaining the
different eigenvalues of $B$ upon measurement when the system state is
initially $A_i$, we must appeal to the probability amplitudes $\chi
(A_i,B_j) $ for the answers. In answering that question, the probability
amplitude $\Psi (A_i,x)$ does not figure, unless we do not have a ready-made
expression for $\chi (A_i,B_j)$ and we seek to obtain it by means of the
expansion Eq. (\ref{fo4}). Since measurements from the states of the
observable $A$ to the observable $B$ do not involve the ''superposition''
state $\Psi (A_i,x)$ but are instead described by the probability amplitude $%
\chi (A_i,B_j),$ the issue of wave function collapse simply does not play a
role in the measurement. The measurement of $B$ throws the system from an
eigenstate of $A$ to an eigenstate of $B$. That is all there is to it. No
reduction of the wave function occurs.

We remark that the cat paradox and wave function collapse are much more
pervasive and ubiquitous in quantum theory than one would at first imagine.
Even the seemingly-innocuous probability amplitude $\chi (A_i,B_j)$ could
easily engender the cat paradox and become susceptible to wave function
reduction if we decided that its simple form is not satisfactory and we
should express the probability amplitude as an expansion. Using the third
observable of the system we obtain 
\begin{equation}
\chi (A_i,B_j)=\sum\limits_n\Psi (A_i,C_n)\xi (C_n,B_j)  \label{si6}
\end{equation}
where $\xi (C_n,B_j)$ are the probability amplitudes for measurements from
states of the observable $C$ to states of the observable $B$. In view of
property Eq. (\ref{th3}) of the probability amplitudes, we have 
\begin{equation}
\xi (C_n,B_j)=\phi ^{*}(C_n,B_j)  \label{se7}
\end{equation}
We can in fact express this as 
\begin{equation}
\chi (A_i,B_j)=\sum\limits_n\Psi (A_i,C_n)\phi ^{*}(C_n,B_j)  \label{ei8}
\end{equation}
We see that we again have a superposition state if we choose to think that
the label $n$ is the state label. Hence, we have a Schr\"odinger cat state.
Moreover, if we choose to analyse the problem of measuring the observable $C$
by means of this probability amplitude instead of by the appropriate
probability amplitudes $\Psi (A_i,C_n)$, we are faced with wave function
collapse, since in that case any value $C_n$ can result from the measurement.

We can go further with this argument. Since any probability amplitude in
quantum mechanics can be expressed as an expansion, we conclude that in fact
all states are superposition states. In particular, even probability
amplitudes representing repeat measurements are superposition states!

A simple example suffices to demonstrate this. Suppose that the energy of a
system is measured and the value $E_k$ is obtained. The measurement has left
the system in the state $\psi _k(x).$ If we measure the energy again, the
same value is obtained with certainty and so the probability amplitude for
this is 
\begin{equation}
\zeta (E_k,E_k)=e^{i\alpha }  \label{ni9}
\end{equation}
where $\alpha $ is a real constant. But this can be expressed as 
\begin{equation}
\zeta (E_k,E_k)=\sum\limits_i\eta (E_k,B_j)\varphi (B_j,E_k)  \label{te10}
\end{equation}
We note in passing that since $\eta (E_k,B_j)=\varphi ^{*}(B_j,E_k)$, Eq. (%
\ref{te10}) is essentially the closure relation.

The expansion Eq. (\ref{te10}) tells us that even the probability amplitude $%
e^{i\alpha }$ is a superposition state which therefore contains wave
function reduction and the cat paradox! If it is argued that the expansion
of this trivial probability amplitude in the form Eq. (\ref{te10}) is
gratuitous, then the reply is that in principle the same applies to the
expansion of the probability amplitude $\Psi (A_i,x).$ Where there is an
expression for the probability amplitude $\Psi (A_i,x)$, there is no need
for an expansion, and hence there is no necessary derivation of either the
cat paradox or wave function collapse from the analysis.

Given the foregoing, we can see that the Schr\"odinger cat paradox is
artificial. The moment we stop thinking of the expansion of a probability
amplitude in terms of the states of a third observable as being a limbo
superposition state, the cat paradox is banished. The expansion Eq. (\ref
{on1}) should really be written as

\begin{equation}
\Psi (A_i,C_n)=\chi (A_i,B_{\text{dead}})\psi (B_{\text{dead}},C_n)+\chi
(A_i,B_{\text{alive}})\psi (B_{\text{alive}},C_n)  \label{el11}
\end{equation}
to emphasize that the actual state is described by $A_i$, and that the
observable $B$ whose eigenvalues are $B_{\text{dead}}$ and $B_{\text{alive}}$
appears arbitrarily; the expansion of the primary probability amplitude $%
\Psi (A_i,C_n)$ is optional$.$ In fact, if this system has another
observable $D$ whose eigenvalues are $D_{\text{fat}}$ and $D_{\text{thin}}$,
this very same probability amplitude can be written as

\begin{equation}
\Psi (A_i,C_n)=\xi (A_i,D_{\text{fat}})\mu (D_{\text{fat}},C_n)+\xi (A_i,D_{%
\text{thin}})\mu (D_{\text{thin}},C_n)  \label{tw12}
\end{equation}
In both cases, the system state is given by $A_i$. If we turn our attention
to the issue of whether the value of the observable $B$ will turn out to be $%
B_{\text{dead}}$ or $B_{\text{alive}}$ upon measurement, we should base our
deliberations on the probability amplitudes $\chi (A_i,B_{\text{dead}})$ and 
$\chi (A_i,B_{\text{alive}}),$ and not on the probability amplitude $\Psi
(A_i,C_n).$

To amplify on the issue of gratuitous expansion of a probability amplitude,
consider the case of probability amplitudes connecting initial states of
given energy to final states of specified position. These probability
amplitudes are just the solutions of the time-independent Schr\"odinger
equation. Since a great deal of quantum mechanics consists in solving this
equation, these functions are known for many cases. We normally think of the
eigenfunctions of the Schr\"odinger equation as expressing pure states. But
if we decide to expand one of them using a third observable of the system,
we end up with a ''superposition state''. A standard such expansion is 
\begin{equation}
\psi _{E_n}(x)=\int e^{ikx}\Phi _{E_n}(p)dp  \label{th13}
\end{equation}
where $\Phi _{E_n}(p)$ are the momentum space eigenfunctions for the system.
If we apply the reasoning that leads to the cat paradox, we must say that
the state of the system is determined by the quantum numbers for the
momentum. Thus, we have the cat paradox and the problem of wave function
reduction arising if we address the problem of measuring the momentum of the
system with reference to this state rather than to the correct probability
amplitude $\Phi _{E_n}(p).$ But of course in this case, it is accepted that
the state of the system is described by the particular value of $E$ which
obtains. In fact, we can clarify this expression by first setting $\psi
_{E_n}(x)=\psi (E_n,x)$ so as to show that this is the probability amplitude
for obtaining specified values of the position when the system state is
initially characterised by the energy $E_n$. We next write $\Phi
_{E_n}(p)=\Phi (E_n,p),$ because this is the probability amplitude for
obtaining specified values of the momentum when the system state is
initially characterised by the energy $E_n.$ We then write $e^{ikx}=\chi
(p,x);$ this is the probability amplitude for obtaining specified values of
the position when the system is initially in the state of momentum $p$. We
then have 
\begin{equation}
\psi (E_n,x)=\int \Phi (E_n,p)\chi (p,x)dp  \label{fo14}
\end{equation}

Here we see that the system state is characterised only by $E_n$, and not by
the label $p$ which occurs only because of the optional decision to use the
eigenfunctions of $p$ to expand the function $\psi (E_n,x)$ (whose form is
already known). There is neither wave function reduction nor cat paradox
proceeding from this state. Where the form of the probability amplitude $%
\psi (E_n,x)$ is known, the expansion Eq. (\ref{fo14}) is not needed, but of
course, this is not to deny the important role that this expansion plays in
the formal development of quantum theory.

Incidentally, we may use this discussion to reinforce the interpretation of
eigenfunctions. The connection between the states involved and the
parameters and variables appearing in an eigenvalue equation of which the
eigenfunctions are solutions is well illustrated when we remind ourselves of
the eigenvalue equations which the functions under discussion satisfy.

The function $\psi (E_n,x)$ is the solution of the time-independent
Schr\"odinger equation 
\begin{equation}
H(x)\psi (E_n,x)=E_n\psi (E_n,x)  \label{fi15}
\end{equation}
The function $\Phi (E_n,p)$ is the solution of the time-independent
Schr\"odinger equation 
\begin{equation}
H(p)\Phi (E_n,p)=E_n\Phi (E_n,p)  \label{si16}
\end{equation}
while the function $e^{ikx}=\chi (p,x)$ is a solution of the eigenvalue
equation 
\begin{equation}
-i\hbar \frac \partial {\partial x}\chi (p,x)=p\chi (p,x)  \label{se17}
\end{equation}

In each case, the eigenvalue corresponds to the initial state, while the
variable in terms of which the differential operator is cast defines the
final state.

\section{Discussion}

In this paper, we have argued that the cat paradox and the problem of wave
function reduction are merely artifices of a misunderstanding of quantum
theory. We have explained how they can be resolved, but our analysis has
produced several questions that deserve to be touched upon.

One of the most profound consequences of the cat paradox is the argument
that certain quantities have no objective existence before they are
measured. According to the present approach, the probability amplitudes of
quantum theory do not warrant this conclusion. We may best interpret them as
giving probabilities for obtaining values of one observable upon measurement
after a previous measurement had thrown the system in question into an
eigenstate of one of its observables. The question whether or not the
quantities being measured pre-exist is not of over-riding importance. But
one thing is certain. The fact that for a given system only certain
quantities can be measured shows that something or other inheres in the
system and makes possible only the measurement of those quantities.
Something therefore pre-exists that relates directly to those measurable
properties of the system. Thus, for example, the fact that no spin
projection can be measured for a particle of spin zero shows that something
pre-exists that precludes the possibility of obtaining non-zero values of
the spin projection.

We emphasize that, in our opinion, the cat paradox and the problem of wave
function reduction are primarily a consequence of confusion and of
mislabelling of quantum quantities. One reason for this, already pointed out
by Land\'e[4], is the pre-eminent role given in quantum theory to solutions
of the Schr\"odinger equation at the expense of other probability
amplitudes. If solutions of the Schr\"odinger equation in coordinate
(momentum) space are interpreted as probability amplitudes for obtaining
eigenvalues of position (momentum) when the system is in a given energy
eigenstate, then it is clear that for that system other probability
amplitudes can be defined. If we do not appreciate this fact sufficiently,
we are tempted to address all problems of measurement to the solutions of
the Schr\"odinger equation, when we should sometimes seek for the answers
elsewhere.

The solutions of the Schr\"odinger equation are only one possible set among
those that can be used to define energy states. Any other probability
amplitudes which have energy states for their initial states, but variables
other than the position for the final observables, are also energy states.
We acknowledge that, in general, the solutions of the Schr\"odinger equation
in coordinate space tend to be the most useful since in many cases we find
ourselves interested in working out expectation values and other properties
of quantities which are most conveniently expressed in terms of the
coordinates. But this should not detract from the fact that other
probability amplitudes corresponding to the energy as defining the initial
eigenvalue also constitute energy states of a system.

.We conclude the paper by remarking that this approach to quantum mechanics,
which derives much of its utility from the principle that all probability
amplitudes should be characterised by a well-defined initial state and a
well-defined final state, is very powerful. For example, it has enabled us
to make useful generalizations of angular momentum theory[9-15]. It very
likely could be used to clarify more of the conceptual problems of quantum
mechanics.

\section{References}

1. Schr\"odinger, E., Naturwissenschaften {\bf 23}, 807 (1935).

2. Heisenberg, W., ''The Physical Principles of Quantum Theory'' (Chicago
University Press, 1930).

3. Land\'e, A., ''From Dualism To Unity in Quantum Physics'' (Cambridge
University Press, 1960).

4. Land\'e, A., ''New Foundations of Quantum Mechanics'', (Cambridge
University Press, 1965).

5. Land\'e, A., ''Foundations of Quantum Theory'', (Yale University Press,
1955).

6. Land\'e, A., ''Quantum Mechanics in a New Key'', (Exposition Press, 1973).

7. Mweene, H. V., {\it Proposed Differential Equation for Spin 1/2}, to be
published in Physica Scripta.

8 . Mweene, H. V., {\it Generalized Spherical Harmonics}, quant-ph/0211135

9. Mweene, H. V., {\it Derivation of Spin Vectors and Operators From First
Principles}, quant-ph/9905012.

10. Mweene, H. V., {\it Generalised Spin-1/2 Operators And Their Eigenvectors%
}, quant-ph/9906002.

11. Mweene, H. V., {\it Vectors and Operators For Spin 1 Derived From First
Principles}, quant-ph/9906043.

12. Mweene, H. V., {\it Alternative Forms of Generalized Vectors and
Operators for Spin 1/2}, quant-ph/9907031.

13. Mweene, H. V., {\it Spin Description and Calculations in the Land\'e
Interpretation of Quantum Mechanics}, quant-ph/9907033.

14. Mweene, H. V., {\it New Treatment of Systems of Compounded Angular
Momentum}, quant-ph/9907082.

15. Mweene, H. V., {\it Derivation of Standard Treatment of Spin Addition
From Probability Amplitudes}, quant-ph/0003056

\end{document}